\documentclass[hyper]{JHEP} 

\usepackage{epsfig}  %%%%%%%%%%%%%%%%%%%%%%%%%%%%%%%%%%%%%%%%%%%%%%%%%%%%%%%%%%%%%%%%%%%%%%%%%%%% %%%%%%%%%%%% Options: preprint* published, (no)hyper*, paper, draft, %%%%%%% %%%%%%%%%%%%          a4paper*, letterpaper, legalpaper, executivepaper,%%%% %%%%%%%%%%%%          11pt, 12pt*, oneside*, twoside %%%%%%%%%%%%%%%%%%%%%%% %%%%%%%%%%%%%%%%%%%%%%%%%%%%%%%%%%%%%%%%%%%%%%%%%%%%%%%%% *=default %%%%%%%% %%%%%%%%%%%% \title{...} %%%%%%%%%%%%%%%%%%%%%%%%%%%%%%%%%%%%%%%%%%%%%%%%%%% %%%%%%%%%%%% \author{...\\...} %%%%%%%%%%%%%%%%%%%%%%%% \email{...} %%%%%%%% %%%%%%%%%%%% \author{...\thanks{...}\\...} %%%%%%%%%%%%%%%%%%%%%%%%%%%%%%%%% %%%%%%%%%%%% \abstract{...} %%%%%%%%%%%%%%%%%%%%%%%%%%%%%%%%%%%%%%%%%%%%%%%% %%%%%%%%%%%% \keywords{...} %%%%%%%%%%%%%%%%%%%%%%%%%%%%%%%%%%%%%%%%%%%%%%%% %%%%%%%%%%%% \preprint{...} %% or \received{...} \accepted{...} \JHEP{...} % %%%%%%%%%%%% \dedicated{...} %%%%%%%%%%%%%%%%%%%%%%%%%%%%%%%%%%%%%%%%%%%%%%% %%%%%%%%%%%%%%%%%%%%%%%%%%%%%%%%%%%%%%%%%%%%%%%%%%%%%%%%%%%%%%%%%%%%%%%%%%%% %%%%%%%%%%%% \aknowledgments %%%%%%%%%%%%%%%%%%%%%%%%%%%%%%%%%%%%%%%%%%%%%%% %%%%%%%%%%%%%%%%%%%%%%%%%%%%%%%%%%%%%%%%%%%%%%%%%%%%%%%%%%%%%%%%%%%%%%%%%%%% %%%%%%%%%%%% -- No pagestyle formatting. %%%%%%%%%%%%%%%%%%%%%%%%%%%%%%%%%%% %%%%%%%%%%%% -- No size formatting. %%%%%%%%%%%%%%%%%%%%%%%%%%%%%%%%%%%%%%%% %%%%%%%%%%%% Your definitions: %%%%%%%%%%% MINE :) %%%%%%%%%%%%%%%%%%%%%%%%% %   ... 								   % \newcommand{\ttbs}{\char'134}           % \backslash for \tt (Nucl.Phys. :)% \newcommand\fverb{\setbox\pippobox=\hbox\bgroup\verb} \newcommand\fverbdo{\egroup\medskip\noindent% 			\fbox{\unhbox\pippobox}\ } \newcommand\fverbit{\egroup\item[\fbox{\unhbox\pippobox}]} \newbox\pippobox %   ...                                                                    % %%%%%%%%%%%%%%%%%%%%%%%%%%%%%%%%%%%%%%%%%%%%%%%%%
\title{Time Dependent Solution  in Open Bosonic
String Field Theory}
\author{ J. Kluso\v{n}
\footnote{On leave from Masaryk University, Brno}\\
Institute of Theoretical Physics, University of Stockholm, SCFAB\\
SE- 106 91 Stockholm, Sweden \\
and \\
Institutionen f\"or teoretisk fysik\\
BOX 803, SE- 751 08 
Uppsala, Sweden \\
E-mail: \email{josef.kluson@teorfys.uu.se}}   \preprint{\hepth{0208028}}  			  	
 \abstract{In this paper we present  time dependent
solution of the open
bosonic string field theory describing
the motion of the tachyon on unstable
D-brane.}

\def\ket #1{\left|#1\right>}

\begin{document}
%%%%%%%%%%%%%%%%%%%%%
%%%%Introduction %%%%%%%%%
%%%%%%%%%%%%%%%%%%%%
\section{Introduction}\label{first}
Time dependent solutions of the open string field
theory  have been studied very intensively
in the last few months.   In particular, the solution
describing rolling of the tachyon field on non-BPS 
D-brane or D-brane anti-D-brane pair away from the
unstable minimum of the potential to the stable one 
has been studied very intensively in series of 
remarkable papers by A. Sen 
\cite{SenT1,SenT2,SenT3}. This problem has been also discussed
in \cite{Sugimoto,Minahan} and very recently in two
papers \cite{SenT4,Zwiebach2}.

In this paper we will  study this problem  from the
point of view of the open bosonic string field theory (SFT)
\cite{WittenSFT} (For review and extensive list
of references, see \cite{Ohmori,DeSmet,BelovR,Siegel}.).
 We present an exact time dependent  solution
of  SFT that  has many properties that we should expect from
the rolling tachyon solution. In particular, we will show that
in the asymptotic future there are not any 
states  
corresponding to the open string fluctuation modes. We will
also show that this solution is characterised by initial position
and velocity of the tachyon field with agreement with
\cite{SenT1,SenT2,SenT3,SenT4,Zwiebach2}.

In our calculation we will proceed as follows. 
We will presume that there is
 solution of the SFT equation of motion that
can be written using string field identity field \cite{Horowitz:dt}
and any operator of the ghost number one that acts on it.
 This presumption is crucial for our analysis since then
we can easily find an exact solution of the SFT equation of motion and
the form of the modified BRST operator. The similar approach for 
searching of the exact solutions of SFT has been used  recently in
\cite{Takahashi1,Takahashi2,Takahashi3,Kluson:2002kk,
Kluson:2002ex,Kluson:2002hr}. 
Success of this approach  strongly support our conjecture
 that
the identity field $\mathcal{I}$ is the fundamental object in the string field
theory and deserve further study. In particular, this string field theory
identity field has been studied recently in many interesting papers
\cite{Ellwood:2001ig,Matsuo,Kishimoto,Schnabl,Kishimoto:2001de}.

This paper is organised as follows. In the next section 
(\ref{second}) we perform explicit calculation and we find time dependent
solution of the SFT. In (\ref{third}) section we show that there
are not  any plane-wave modes that propagate about the new background
given by the rolling tachyon solution. And finally, in (\ref{fourth}) we
outline our results and suggest further open questions and problems. 

%%%%%%%%%%%%%%%%%%%%%%%%%%%%%%%%%%%%%%%%%%%%%%%%%

%%%%Sen tachyon solution %%%%%%%%%%%%%%%%%%%%%%%%

%%%%%%%%%%%%%%%%%%%%%%%%%%%%%%%%%%%%%%%%%%%%%%%%%
\section{Calculation}\label{second}
In this section we will solve the string field theory equation of motion.
We show that there is an exact time dependent 
solution   that describes rolling of the tachyon 
from its initial value $\lambda $ 
 to the minimum of the potential in the limit $x^0\rightarrow 
\infty $.

To begin with, we firstly review basic facts about
bosonic string field theory, following
mainly \cite{Ohmori}. Gauge invariant
string field theory is described with the
full Hilbert space of the first quantized open
string including $b,c $ ghost fields subject
to the condition that the states must
carry ghost number one, where $b$ has ghost
number $-1$, $c$ has ghost number $1$ and
$SL(2,C)$ invariant vacuum $\ket{0}$
carries ghost number $0$. We denote
$\mathcal{H}$ the subspace of the full Hilbert
space carrying ghost number $1$.  Any state
in $\mathcal{H}$ will be denoted as $\ket{\Phi}$
and corresponding vertex operator $\Phi$
is the  vertex operator that  creates state $\ket{\Phi}$
out of the vacuum state $\ket{0}$
\begin{equation}
\ket{\Phi}=\Phi \ket{0} \ .
\end{equation}
Since we are dealing with open string theory,
the vertex operators should be put on the boundary
of the world-sheet. In the following we will also use double
cover of the upper half  plane and then we will consider
holomorphic fields only.  The string field theory action
is given \cite{WittenSFT} 
\begin{equation}\label{actionCFT}
S=-\frac{1}{g_0^2}\left(
\frac{1}{2\alpha'}\left<I\circ \Phi (0) Q_B \Phi (0)\right>+
\frac{1}{3}\left<f_1\circ \Phi(0) f_2\circ \Phi(0)
f_3\circ \Phi(0)\right>\right) \ ,
\end{equation}
where $g_0$ is open string coupling constant, $Q_B$
is BRST operator and $<>$ denotes correlation function
in the combined matter ghost conformal field theory.
$I,f_1, f_2, f_3$ are conformal mapping  exact
form of which is reviewed in \cite{Ohmori} and $f_i\circ\Phi(0)$ denotes
the conformal transformation of $\Phi(0)$ by $f_i$. For
example, for $\Phi$ a primary field of dimension $h$,
then $f_i\circ \Phi(0)=(f'_i(0))^h\Phi(f_i(0))$. 
In the abstract language pioneered in \cite{WittenSFT}, the
 open string field theory action (\ref{actionCFT})
has a form (For more details, see
\cite{Ohmori,DeSmet,BelovR,Siegel}.)
\begin{equation}\label{actionW}
S=-\frac{1}{g_0^2}\left(
\frac{1}{2\alpha'}\int \Phi \star Q\Phi
+\frac{1}{3}\int \Phi\star\Phi\star\Phi\right) \ .
\end{equation}
Let us presume that the solution of the string field theory
equation of motion 
\begin{equation}\label{eq}
\frac{1}{\alpha'}Q\Phi_0+\Phi_0\star \Phi_0=0  \ , 
\end{equation}
can be written as
\begin{equation}\label{soT}
\ \Phi_0=\mathcal{T}_L(\mathcal{I}) \  ,
\end{equation}
where $\mathcal{I}$ is string field algebra $\star $ identity
field \cite{Horowitz:dt} that obeys
\begin{equation}
\mathcal{I}\star X=X\star \mathcal{I}=X \ ,
\end{equation}
for any string field $X$.
The ghost number one operator $\mathcal{T}$ is defined as
\begin{equation}\label{def}
\mathcal{T}(\Phi(w))=\frac{1}{2\pi i}\oint_C dz t(z)\Phi(w) \ ,
\end{equation}
where the integration contour $C$ encircles the point $w$  of the 
complex plane counterclockwise and we use the convention
that $t(z)$ is holomorphic field  defined over the whole complex
plane via the doubling trick. In upper expression  $
\Phi(w)$ is any CFT operator. And finally, the subscript $L$ 
in  (\ref{soT}) means that
the integration contour corresponds to the integration 
over left side of the string. In the complex plane, this contour can
be parameterised as $z=-e^{-i\sigma+\tau} , \sigma \in (\pi/2,-\pi/2) $
for $\mathcal{T}_L$ acting on $\ket{\phi}=\phi(0)\ket{0}$.
 For $\mathcal{T}_L$ acting
on any wedge state $\ket{n}$ the situation is slightly more complicated,
for precise discussion of these problems, see very nice papers 
\cite{Kishimoto:2001de,Schnabl}. For our purpose the definition
(\ref{def}) will be sufficient. 

  Using results given in
\cite{Horowitz:dt} and recently discussed in 
\cite{Kishimoto:2001de} we can easily show that 
the field (\ref{soT}) obeys 
\begin{equation}
\Phi_0\star \Phi_0=
\mathcal{T}_L(\mathcal{I})\star \mathcal{T}_L(\mathcal{I})=-
\mathcal{T}_L(\mathcal{I})
\star \mathcal{T}_R(\mathcal{I})=
-\mathcal{T}_L(\mathcal{T}_L(\mathcal{I}))=-
\frac{1}{2}\left\{\mathcal{T}_L,
\mathcal{T}_L\right\}(\mathcal{I}) \ .
\end{equation}
In the same way we get
\begin{eqnarray}\label{comm}
Q\Phi_0=(Q_R+Q_L)
(\mathcal{T}_L(\mathcal{I}))=
Q_L(\mathcal{T}_L(\mathcal{I}))-
\mathcal{T}_L(Q_R(\mathcal{I}))=
\left\{Q,\mathcal{T}\right\}_L(\mathcal{I})  \ , \nonumber \\
\end{eqnarray}
where we have used the fact that $Q_R$ anticommutes
with $\mathcal{T}_L$ and  the relation 
\cite{Horowitz:dt} 
\begin{equation}
Q_L(\mathcal{I})=-Q_R(\mathcal{I})  \ .
\end{equation}
In these formulas the subscript $R$ means the integration
over right side of the string. And finally, we use the notation
\begin{equation}
[X_L,Y_L]=[X,Y]_L \ .
\end{equation}
Let us presume that the solution of (\ref{eq}) can
be written as
\begin{equation}\label{T}
\mathcal{T}_L(\mathcal{I})=
T_L(\mathcal{I})+\delta T_L(\mathcal{I}) \ ,
Q(T_L(\mathcal{I}))=\left\{Q,T\right\}_L(\mathcal{I})=0 \ 
\end{equation}
and  $\delta T$ obeys
\begin{equation}
\left\{Q,\delta T\right\}_L-\frac{1}{2}
\left\{T,T\right\}_L=0 \ , \left\{\delta T,
\delta T\right\}_L=0 \ ,
\left\{\delta \mathcal{T},T\right\}=0 \ .
\end{equation}
Then it is clear that  (\ref{T})
is 
exact solution of the string field theory equation of motion.
 
Let us consider  $T$ in the form
\begin{equation}
T=\frac{1}{2\pi i}\oint dw t(X^0(w))c(w)  \ ,
t(X^0(w))=\sum_n a_n X^0(w)^n  \ ,
\end{equation}
where the normal
ordering between $X^0$ is understood. 
In the following we use the convention
\cite{Ohmori,Polchinski}
\begin{eqnarray}
T_m(z)=-\frac{1}{\alpha'}\partial_z X^{\mu}(z)
\partial_z X^{\nu}\eta_{\mu\nu} 
 \ , \nonumber \\
 X^{\mu}(z)X^{\nu}(w)\sim
-\frac{\alpha'}{2}\eta^{\mu\nu}
\ln (z-w)  \ , \nonumber \\
 c(z)b(w)\sim -\frac{1}{z-w} \ ,  \nonumber \\
Q=\frac{1}{2\pi i}\oint dz c(z)
\left[T_m(z)+\frac{1}{2}T_{gh}(z)\right] \ ,
\nonumber \\
T_{gh}(z)=-2b(z)\partial c(z)-\partial (b(z)) c(z)
 \ ,  \nonumber \\
T_{gh}(z)c(w)\sim
\frac{-1}{(z-w)^2}c(w)+\frac{\partial c(w)}{z-w} \ .
\nonumber \\
\end{eqnarray}
Then we obtain from (\ref{T})
\begin{eqnarray}\label{eqm1}
0=\left\{Q,T\right\}_L = \nonumber \\
=\left\{\frac{1}{2\pi i}\int_{C_1}
 dz c(z)\left[T_m(z)
+\frac{1}{2}T_{ghost}(z)\right], 
\frac{1}{2\pi i}\int_{C_2} dw t(X^0(w))c(w)\right\}
=\nonumber \\
=\frac{1}{2\pi i}\int_{C_1} dw \partial c(w)
c(w)\left[\frac{\alpha'}{4}t''(X^0(w))-t(X^0(w))
\right] \ , \nonumber \\
\end{eqnarray}
where $C_1,C_2$ are integration contours in the 
complex plane corresponding to the integration
over left side of the string. Finally, we have 
defined $t''(X^0(w))=\frac{d^2 t(x)}{d^2x}$.
From the requirement of the vanishing
(\ref{eqm1})  we immediately get two
 linearly
 independent 
solutions 
\begin{equation}
t_1(w)=e^{\frac{2}{\sqrt{\alpha'}}
X^0(w)} , \ t_2(w)=
e^{-\frac{2}{\sqrt{\alpha'}}
X^0(w)} \ .
\end{equation}
Now we can ask the question whether the linear combination
of these two solutions 
\begin{equation}\label{sollin}
T=AT_1+BT_2 
\end{equation}
is exact solution of the SFT equation of motion. 
It is clear that this operator obeys
\begin{equation}
\left\{Q,T\right\}_L=0 \ 
\end{equation}
and hence it is solution of the SFT equation of motion in
the linearised approximation
  \cite{SenT1,SenT2}.
 In order to determine whether
 the string field $T_L(\mathcal{I})$ is  solution of
(\ref{eq})  we must calculate
 following expression
\begin{equation}\label{comm2}
\frac{1}{2}
\left\{T_L,T_L\right\}=
\frac{1}{2}\left\{\frac{1}{2\pi i}\int_{C_1}dz c(z)t(X^0(z)),
\frac{1}{2\pi i}\int_{C_2}dw c(w)t(X^0(w))\right\} \ ,
\end{equation}
where $C_1,C_2$ are the same integration contours
 as in the previous example.  
Now we rewrite (\ref{comm2}) as
\begin{equation}\label{comm2a}
\frac{1}{4\pi i}\int_{C_1}dw 
\left[\frac{1}{2\pi i} \oint_{C'}c(z)t(X^0(z))t(X^0(w))
\right]c(w) \ ,
\end{equation}
where the integration contour
$C'$ encircles the point $w$ counterclockwise.
From upper expression it is clear that we must determine 
Operator product expansion (OPE)
between $t(z),t(w)$. Using the well known formula
 \cite{Polchinski} for the calculation of the OPE between two
normal ordered operators
$:\mathcal{F}:,:\mathcal{G}:$ 
\footnote{In the following we introduce the symbol $:F :$
for the  normal ordered operator $F$.}
\begin{equation}
:\mathcal{F}(z)::\mathcal{G}(w):
=\exp\left(-\frac{\alpha'}{2}\int dz_1dz_2 \ln (z_1-z_2)
\eta^{\mu\nu}\frac{\delta_{\mathcal{F}}}
{\delta X^{\mu}(z_1)}
\frac{\delta_{\mathcal{G}}}
{\delta X^{\nu}(z_2)}\right)
:\mathcal{F}(z)\mathcal{G}(w):  \ 
\end{equation}
we get
\begin{eqnarray}\label{OPE11}
t_1(z)t_1(w)=:e^{\frac{2}{\sqrt{\alpha'}}
X^0(z)}:
:e^{\frac{2}{\sqrt{\alpha'}}X^0(w)}:=
\exp \left(-\frac{\alpha'}{2}
\int dz_1dz_2 \ln (z_1-z_2)
\frac{2}{\sqrt{\alpha'}}\delta^0_{\mu}
\delta (z-z_1)\times \right.
\nonumber \\ \left. \times
\frac{2}{\sqrt{\alpha'}}\delta^0_{\nu}
\delta (w-z_2)\delta^0_{\nu}
\eta^{\mu\nu}\right):
e^{\frac{2}{\sqrt{\alpha'}}
X^0(z)}
e^{\frac{2}{\sqrt{\alpha'}}X^0(w)}:
=(z-w)^{2}
:e^{\frac{2}{\sqrt{\alpha'}}
X^0(z)}
e^{\frac{2}{\sqrt{\alpha'}}
X^0(w)}:
    \ , \nonumber \\
t_2(z)t_2(w)=
:e^{-\frac{2}{\sqrt{\alpha'}}
X^0(z)}:
:e^{-\frac{2}{\sqrt{\alpha'}}X^0(w)}:=
(z-w)^{2}
:e^{-\frac{2}{\sqrt{\alpha'}}
X^0(z)}
e^{-\frac{2}{\sqrt{\alpha'}}
X^0(w)}:
    \ , \nonumber \\
\end{eqnarray}
where we have used
\begin{eqnarray}
\frac{\delta}{\delta
X^{\mu}(z_1)}t_1(z)=
\frac{\delta}{\delta
X^{\mu}(z_1)}e^{\frac{2}{\sqrt{\alpha'}}
X^0(z)}=
\delta (z-z_1)\frac{2}{\sqrt{\alpha'}}
\delta^0_{\mu}
e^{\frac{2}{\sqrt{\alpha'}}
X^0(z)} \ , \nonumber \\
\frac{\delta}{\delta
X^{\nu}(z_2)}t_2(w)=
\frac{\delta}{\delta
X^{\nu}(z_2)}e^{-\frac{2}{
\sqrt{\alpha'}}X^0(w)}
=-\delta (w-z_2)\frac{2}{\sqrt{
\alpha'}}\delta^0_{\nu}
e^{-\frac{2}{\sqrt{\alpha'}}X^0(w)} \ .\nonumber \\
\end{eqnarray}
We see from (\ref{OPE11}) 
that the OPE between 
$t_1(z)t_1(w) \ ,t_2(z)t_2(w) $ are non-singular and 
consequently
\begin{eqnarray}\label{comTT}
\left\{T_1,T_1\right\}_L=0 \ , \nonumber \\
\left\{T_2,T_2\right\}_L=0 \ . \nonumber \\
\end{eqnarray}
We can also show that 
\begin{eqnarray}\label{OPE12}
t_1(z)t_2(w)=:e^{\frac{2}{\sqrt{\alpha'}}
X^0(z)}:
:e^{-\frac{2}{\sqrt{\alpha'}}X^0(w)}:=\nonumber \\
=(z-w)^{-2}
:e^{\frac{2}{\sqrt{\alpha'}}
X^0(z)}e^{-\frac{2}{\sqrt{\alpha'}}X^0(w)}:
=\frac{1}{(z-w)^2}:e^{\frac{2}{\sqrt{\alpha'}}
X^0(w)}e^{-\frac{2}{\sqrt{\alpha'}}X^0(w)}:+ \nonumber \\
+
\frac{1}{(z-w)^2}\frac{2}{\sqrt{\alpha'}}:
\partial_w X^0(w)(z-w)e^{\frac{2}{\sqrt{\alpha'}}
X^0(w)}e^{-\frac{2}{\sqrt{\alpha'}}X^0(w)}:+\dots= \nonumber \\
=\frac{1}{(z-w)^2}+
\frac{1}{z-w}\frac{2}{\sqrt{\alpha'}}\partial_w X^0(w)+\dots \ . \nonumber \\
\end{eqnarray}
Using  (\ref{OPE11}),(\ref{comTT}),(\ref{OPE12}) we get
\begin{eqnarray}\label{TT12}
\frac{1}{2}
\left\{T_L,T_L\right\}=
AB\left\{T_{1L},T_{2L}\right\}=AB
\frac{1}{2\pi i}\int_{C_1}dw \left[
\frac{1}{2\pi i} \oint_{C'}c(z)t_1(X^0(z))t_2(X^0(w))\right]c(w)
=\nonumber \\
=AB\frac{1}{2\pi i}\int_{C_1}dw \left[
\frac{1}{2\pi i}\int_{C'}dz \left(
\frac{1}{(z-w)^2}+\frac{2}{\sqrt{\alpha'}}
\frac{1}{z-w}\partial_wX^0(w)\right)c(z)\right]c(w)=\nonumber \\
=AB
\frac{1}{2\pi i}\int_{C_1}dw \partial_w c(w)c(w)+
c(w)\frac{2}{\sqrt{\alpha'}}\partial_wX^0(w)c(w)=AB
\frac{1}{2\pi i}\int_{C_1}dw \partial_w c(w) c(w)
\neq 0 
 \  \nonumber \\
\end{eqnarray}
and hence
 the linear combination
$AT_1+BT_2$  is not an exact solution of 
the string field theory equation of motion.
As in (\ref{T}) we introduce following term
\begin{equation}\label{deltaT}
\delta \mathcal{T}=-AB\alpha'
\frac{1}{2\pi i}\int_{C_1}dz c(z) \ . 
\end{equation}
From the upper expression  we immediately obtain
\begin{equation}
\left\{\delta \mathcal{T},\delta \mathcal{T}\right\}_L=0 \ ,
\left\{\delta \mathcal{T},T_{1,2}\right\}=0 \ . 
\end{equation}
 We can also easily see that
\begin{eqnarray}\label{deltaT2}
\frac{1}{\alpha'}\left\{
Q,\delta \mathcal{T}\right\}_L=
\left\{\frac{1}{4\pi i}\int_{C_1}dz c(z)T_g(z),
-\frac{AB}{2\pi i}\int_{C_2}dwc(w)\right\}=
\nonumber \\
=-\frac{AB}{4\pi i}\int_{C_1}dw \left[
\frac{1}{2\pi i}\oint_{C'}c(z)\left(
-\frac{1}{(z-w)^2}c(w)+\frac{1}{z-w}\partial_wc(w)\right)
\right]=\nonumber \\
=-\frac{AB}{4\pi i}\int_{C_1}dw\left\{
-\frac{\partial}{\partial_w}\left[\oint_{C'}dzc(z)\frac{1}{z-w}
\right]c(w)+
c(w)\partial_wc(w)\right\}=\nonumber \\
=\frac{AB}{2\pi i}\int_{C_1}dw 
\partial_wc(w) c(w) \ . \nonumber \\
\end{eqnarray}
From (\ref{TT12}) and  (\ref{deltaT2}) 
we immediately get
\begin{equation}
\frac{1}{\alpha'}\left\{Q,\delta\mathcal{T}\right\}_L-
\frac{1}{2}\left\{\mathcal{T},\mathcal{T}\right\}_L=0 \ 
\end{equation}
so that 
\begin{equation}\label{exactT}
\Phi_0=\mathcal{T}_L(\mathcal{I}) \ ,
\mathcal{T}=\frac{1}{2\pi i}\oint
dz
c(z)\left(A e^{\frac{2}{\sqrt{\alpha'}}X^0(z)}+
Be^{-\frac{2}{\sqrt{\alpha'}}X^0(z)}-\alpha'AB\right) \ .
\end{equation}
is exact solution of the SFT equation of motion. 

Let us consider  following initial  condition
\cite{SenT1}
\begin{equation}\label{initial}
\mathcal{T}(X^0=0)=\lambda c(z), \ 
 \mathcal{T}'(X^0=0)=0 \ .
\end{equation}
From the second condition we get
\begin{equation}
0=\frac{d \mathcal{T}(X^0)}{dX^0}(X^0=0)=
c(z)(A-B) \Rightarrow
A=B  \ ,
\end{equation}
and when we insert this result into the first
condition in (\ref{initial})
we immediately obtain
\begin{eqnarray}
\mathcal{T}(X^0=0)=c(z)\left(
-\alpha'AB+A+B\right)=c(z)\lambda 
\Rightarrow -\alpha'A^2+2A-\lambda=0 \Rightarrow
\nonumber \\
\Rightarrow
A_1=\frac{1}{\alpha'}\left(1+\sqrt{1-\alpha'\lambda}\right) \ ,
A_2=\frac{1}{\alpha'}\left(1-\sqrt{1-\alpha'\lambda}\right) \ .
\nonumber \\
\end{eqnarray}
For
$\lambda\ll 1/\alpha' $ we get
\begin{equation}
A_1\sim \frac{1}{\alpha'}\left(1+(1-\alpha'\lambda/2)\right)=
\frac{1}{\alpha'}\left(2-\alpha'\lambda/2\right) \ .
A_2\sim \frac{1}{\alpha'}(1-(1-\alpha'
\lambda/2))=\frac{\lambda}{2} \ .
\end{equation}
For $\lambda \rightarrow 0$,  $A_2$ goes
to zero and hence the solution $\Phi_0$ vanishes.
On the other hand for $\lambda \rightarrow 0$ 
$A_1$ goes to $2/\alpha'$. In this case the tachyon field
would start to roll even if its initial value and velocity are zero.
We mean that in the classical string field
 theory there is no reason for
such a behaviour so that we will not consider the first
root $A_1$.

To obtain the string field theory
 action for the fluctuation fields about
the classical solution $\Phi_0$ we insert the general
string field $\Phi=\Phi_0+\Psi $ into (\ref{actionW}).
Then we get the string field theory
action for the string field $\Psi$ which has the same
form as (\ref{actionW}) however the original BRST
operator $Q$ is replaced with the new one
\begin{equation}
Q'(\Psi)=Q(\Psi)+\Phi_0\star \Psi-(-1)
^{|\Psi|}\Psi \star \Phi_0=
Q(\Psi)-\mathcal{T}(\Psi) \ ,
\end{equation}
using \cite{Horowitz:dt,Kishimoto:2001de}
\begin{eqnarray}
\Phi_0\star\Psi=
\mathcal{T}_L(\mathcal{I})\star \Psi=
-\mathcal{I}\star \mathcal{T}_R(\Psi)=-
\mathcal{T}_R(\Psi) \ , \nonumber \\
-(-1)^{|\Psi|}\Psi\star\Phi_0=
(-1)^{|\Psi|}\Psi\star\mathcal{T}_R(\mathcal{I})=
-\mathcal{T}_L(\Psi)\star\mathcal{I}=-\mathcal{T}_L
(\Psi) \ ,
\nonumber \\
\mathcal{T}_L(\mathcal{I})=-\mathcal{T}_R(\mathcal{I}) \ ,
\mathcal{T}(\Psi)=\mathcal{T}_L(\Psi)+
\mathcal{T}_R(\Psi) \  . \nonumber \\
\end{eqnarray}
As a result we obtain the new BRST operator
$Q'$ in the form
\begin{eqnarray}\label{Qshift}
Q'=\frac{1}{2\pi i}\oint dz c(z)
\left[T_m(z)-A e^{\frac{2}{\sqrt{\alpha'}}X^0(z)}-
Be^{-\frac{2}{\sqrt{\alpha'}}X^0(z)}+\alpha'AB
+
\frac{1}{2}T_{gh}(z)\right] =\nonumber \\
=\frac{1}{2\pi i}\oint dz c(z)
\left[T_m(z)-2A\cosh\left(
\frac{2}{\sqrt{\alpha'}}X^0(z)\right)+\alpha'A^2
+
\frac{1}{2}T_{gh}(z)\right] \ . \nonumber \\
\end{eqnarray}
In the next section we will study the fate of  
the fluctuation modes about the solution
 $\Phi_0=\mathcal{T}_L
(\mathcal{I})$. 
%%%%%%%%%%%%%%%%%%%%%%%%%%%%%%%%
%%%%%Absence of fluctuation fields %%%%%%%%%
%%%%%%%%%%%%%%%%%%%%%%%%%%%%%%%%
\section{Absence of the perturbative states in
the asymptotic future}\label{third}
In this section we will analyse  fluctuation modes
about  the solution (\ref{exactT}) 
in the limit $X^0\rightarrow \infty$. According to 
very nice analysis given in  \cite{SenT3},
 there should not 
be any plane-wave perturbative states 
in the limit $X^0\rightarrow \infty $. 
We will argue that the same result holds for the fluctuation
fields about the string field solution $\mathcal{T}_L(
\mathcal{I})$. We hope that this conclusion strongly
supports our conjecture that (\ref{exactT}) really describes
rolling the tachyon in the string field theory.

To begin with, we
propose following form of the fluctuation field. We conjecture
that any fluctuation mode about the solution
$\Phi_0=\mathcal{T}_L(\mathcal{I})$ has the form
\begin{equation}\label{fluctx}
\Psi=F(c^{\dag},b^{\dag},a^{\dag})(\mathcal{T}_L(
\mathcal{I})) \ ,
\end{equation}
where $F(c^{\dag},b^{\dag},a^{\dag})$ is ghost number zero
operator. 
The proposal (\ref{fluctx}) is mainly motivated by recent
works \cite{Hata1,Hata2} where the fluctuation
mode about sliver state in VSFT has been studied. 
Other aspects of the fluctuation modes about sliver state
have been  discussed  in
\cite{Rastelli, Rashkov,Rashkov2,Rashkov3,Okawa,Hata3}. 

In CFT language 
the fluctuation field (\ref{fluctx}) can be written  as
\begin{eqnarray}\label{fluctcft}
\Psi=\frac{1}{2\pi i}
\oint dw  f(X(w),\partial X(w))
\left(\frac{1}{2\pi i}\int_{C_1}dz \mathcal{T}(z)
\mathcal{I}\right)=\nonumber \\
=
\frac{1}{2\pi i}\int_{C_1}dz \left[
\frac{1}{2\pi i}\oint_{C'} dw f(w)\mathcal{T}(z)\right]
(\mathcal{I})=
[F,\mathcal{T}]_L(\mathcal{I}) \ ,
\nonumber \\
\end{eqnarray}
using 
\begin{equation}
F_L(\mathcal{I})=-F_R(\mathcal{I}) \ .
\end{equation}
Now we will try to find spectrum
of the fluctuation modes. 
 By definition,  the
fluctuation modes obey
the  linearised string field theory equation
of motion formulated about the new background
\begin{equation}\label{Qbar}
Q'\left([F,\mathcal{T}]_L(\mathcal{I})\right)=0
\Rightarrow
\left\{Q,[F,\mathcal{T}]\right\}_L-
\left\{\mathcal{T},[F,\mathcal{T}]\right\}_L=0 \ .
\end{equation}
To solve previous equation, we  need following
formulas
\begin{eqnarray}
\left\{Q,[F,\mathcal{T}]\right\}_L-
\left[F,\left\{\mathcal{T},Q\right\}
\right]_L-\left\{\mathcal{T},[Q,F]\right\}_L=0 \Rightarrow \  , \nonumber \\
\Rightarrow \left\{Q,[F,\mathcal{T}]\right\}_L=
\left[F,\left\{\mathcal{T},Q\right\}
\right]_L+\left\{\mathcal{T},[Q,F]\right\}_L=\nonumber \\
=
\left[F,\left\{\delta \mathcal{T},Q\right\}\right]_L+
\left\{\mathcal{T},[Q,F]\right\}_L
=\left\{\mathcal{T},[Q,F]\right\}_L\ , \nonumber \\
\end{eqnarray}
where we have used the fact that $F$ is the ghost number zero 
operator and hence has trivial OPE with any ghost field so
that $[F,\left\{Q,\delta T\right\}]=0$. 

In the same way we get
\begin{eqnarray} 
\left\{\mathcal{T},[F,\mathcal{T}]\right\}_L=
\left\{\delta \mathcal{T}+T_1+T_2,
[F,T_1+T_2]\right\}_L=\nonumber \\
=\left\{T_1,[F,T_2]\right\}+
\left\{T_2,[F,T_1]\right\}_L=\nonumber \\
=\left[F,\left\{T_1, T_2\right\}\right]_L=2
\left[F,\left\{Q,\delta\mathcal{T}\right\}\right]_L=0 \ . \nonumber \\
\end{eqnarray}
Then  (\ref{Qbar}) reduces to
\begin{equation}\label{QF}
0=Q'(F(\mathcal{T}_L(\mathcal{I}))=
\left\{\mathcal{T},[Q,F]\right\}_L
\Rightarrow [Q,F]_L=0 \ .
\end{equation}
We are interested in the limit $X^0\rightarrow \infty $ so 
that
$\mathcal{T}\rightarrow \frac{A}{2\pi i}
\oint dzc(z)e^{\frac{2}{\sqrt{\alpha'}}X^0(z)}$.
Now we would like to show, in the same way as in
\cite{SenT3}, that in the asymptotic future  there are not
any plane-wave  states around the tachyon vacuum. 
Let us start with the operator 
\begin{equation}
F=\frac{1}{2\pi i}\oint dw f(w)=
\frac{1}{2\pi i}\oint dw e^{2ik_{\mu}X^{\mu}(w)} \ .
\end{equation}
Then (\ref{QF}) gives the condition 
\begin{equation}\label{masshell} 
\left(\alpha' k_{i}k_{j}\eta^{ij}
-\alpha'k_0^2\right)=
-k_{\mu}k_{\nu}\eta^{\mu\nu}=0 
\ .
\end{equation}
Note that there is not any contribution from the ghost sector
since $F$ is the ghost number zero operator and hence has
trivial OPE with $T_{ghost}$. According
to (\ref{fluctcft}) we must  calculate following commutator 
\begin{eqnarray}\label{fluct}
[F,\mathcal{T}(X^0\rightarrow \infty)]_L=
\left[\frac{1}{2\pi i}\int_{C_1}dz
e^{2ik_0X^0(z)},\frac{A}{2\pi i}\int_{C_2}
dw c(w) e^{\frac{2}{\sqrt{\alpha'}}X^0(w)}\right]=
\nonumber \\
=\frac{1}{2\pi i}\int_{C_1}dw c(w) \left[
\frac{A}{2\pi i} \oint_{C'}dz   
e^{2ik_0X^0(z)}e^{\frac{2}{\sqrt{\alpha'}}X^0(w)}\right] \ .
\nonumber \\
\end{eqnarray}
In order to obtain nontrivial result 
we should consider $k_0$ in the form
\begin{equation}\label{kcond}
k_{0}=\frac{in}{\sqrt{\alpha'}}  \ , n=1,2\dots 
\end{equation}
so we get
\begin{eqnarray}\label{nOPE}
:e^{-\frac{2n}{\sqrt{\alpha'}}X^0(z)}:
:e^{\frac{2}{\sqrt{\alpha'}}X^0(w)}:=
:\exp\left(\frac{\alpha'}{2}\int dz_1dz_2
\ln (z_1-z_2)\frac{-2n}{\sqrt{\alpha'}}
\delta(z_1-z)\frac{2}{\sqrt{\alpha'}}
\delta (z_2-w)\right)\times \nonumber \\
\times 
:e^{-\frac{2n}{\sqrt{\alpha'}}X^0(z)}
e^{\frac{2}{\sqrt{\alpha'}}X^0(w)}:=
\frac{1}{(z-w)^{2n}}:e^{-\frac{2n}{\sqrt{\alpha'}}X^0(z)}
e^{\frac{2}{\sqrt{\alpha'}}X^0(w)}: \ .
\nonumber \\
\end{eqnarray}
When we insert (\ref{nOPE}) into 
(\ref{fluct}) we obtain well defined  operator
and hence well defined fluctuation mode.
On the other hand, the  condition
(\ref{masshell}) implies
\begin{equation}
k_ik^i=-\frac{1}{\alpha'}n^2 \ , n=1,2,\dots 
\end{equation}
so that
the spatial part of the momentum is imaginary. This result clearly implies
that 
there are not any fluctuation modes that propagate as plane-wave
solutions about tachyon vacuum with 
 agreement with the analysis performed in
\cite{SenT3}.
It is easy to extend this arguments to other perturbative
modes 
\begin{equation}
F=\frac{1}{2\pi i} \oint dw
 O(\partial_wX(w),\dots)e^{ik_{\mu}X^{\mu}(w)} \ ,
\end{equation}
where now the condition $[Q,F]_L=0$
implies
\begin{equation}
-k_0^2+k_ik^i=-\frac{1}{\alpha'}
N \Rightarrow
k_ik^i=-\frac{1}{\alpha'}
(n^2+N) \ .
\end{equation}
As in previous case the requirement of the well defined operator
leads to the condition (\ref{kcond}) 
which again implies that spatial components of the momentum are
imaginary and hence  there are not present any
 plane-wave excitations.
We mean that this result strongly supports our conjecture that
the solution given in the previous section really 
describes rolling tachyon \cite{SenT1,SenT2,SenT3}. 

%%%%%%%%%%%%%%%%%%%%%%%%
%%%%conclusion %%%%%%%%%%%%%
%%%%%%%%%%%%%%%%%%%%%%%
\section{Conclusion}\label{fourth}
In this short note we have found  an exact solution of the open
bosonic string field theory that represents the time dependent
flow to the tachyonic vacuum. We have seen that this solution is
characterised by one parameter $\lambda$ so that despite the non-locality
of the string field theory action, it does admit time dependent solutions
labelled by initial position and velocity of the tachyon field, with complete
agreement with recent papers \cite{SenT1,SenT2,SenT3,SenT4,Zwiebach2}.  
 We have also shown that in the asymptotic future
there are not any physical fluctuations  about  this solution again
with the agreement with \cite{SenT3}. 

There are many problems and questions that deserve further study.
Since we have argued that the solution given in this paper 
corresponds to the flow of the system to the tachyonic vacuum, we should expect
that the string field theory formulated about this solution will have many
common with the vacuum string field theory.
In particular, we believe that there exist such a string field redefinition that
maps the shifted BRST operator (\ref{Qshift}) 
in the limit $X^0\rightarrow \infty $
  to the pure ghost BRST operator  
in VSFT \cite{SenV2,SenV3,SenV4,SenV5}.

We have seen that there are not any perturbative fluctuations corresponding to
the open string modes about the rolling tachyon solution.
On the other hand,  it is possible that there could be solutions of the string
field theory formulated about rolling tachyon background in the limit $X^0\rightarrow
\infty $
that represent closed string excitations or D-branes of various dimensions.
If our presumption
is correct that  these solutions should be related
to the sliver states in VSFT. 
We hope to return to these important and exciting problems in the future. 

{\bf Acknowledgement}
This work is partly supported 
by EU contract HPRN-CT-2000-00122.

 %%%%%%%%%%%%%%%%%%%%%%%%%%%%%%%%%%%%%%
    %%%%%%% Thebibligraphy %%%%%%%%%%%%%%%%%%%%%
    %%%%%%%%%%%%%%%%%%%%%%%%%%%%%%%%%%%%%
    
    \end{document}